\documentclass[%
 reprint,
 amsmath,amssymb,
 aps,
]{revtex4-1}
\usepackage{mathrsfs}
\usepackage{graphicx}
\usepackage{dcolumn}
\usepackage{bm}

\begin{document}

\title{Single photon transfer controlled by excitation phase in a two-atom cavity system}

\author{Chun Xiao Zhou$^{1,2}$}
\author{Rui Zhang$^{1}$}
\author{Miao Di Guo$^{1}$}
\author{S. A. Moiseev$^{1,3,4}$}
\email{samoi@yandex.ru}
\author{Xue Mei Su$^{1}$}
 \email{suxm@jlu.edu.cn}

\affiliation{%
 $^{1}$Key Lab of Coherent Light, Atomic and Molecular Spectroscopy, Ministry of Education; and College of Physics, Jilin University, Changchun 130012, People's Republic of China\\
$^{2}$College of Physics and Electronics, Hunan University of Arts and Science, Changde 415000, People's Republic of China\\
$^{3}$Quantum Center, Kazan National Research Technical University, 10 Karl Marx Street, Kazan 420111, Russia\\
$^{4}$Zavoisky Physical-Technical Institute of the Russian Academy of Sciences, 10/7 Sibirsky Trakt, Kazan, 420029, Russia}%

\date{\today}

\begin{abstract}
  We investigate the quantum interference effects of single photon transfer in two-atom cavity
system caused by external excitation phase. In the proposed system, two identical atoms (with
different positions in the optical cavity) are firstly prepared into a timed state by an external
single photon field. During the excitation, the atoms grasp different phases which depend on
the spatial positions of the atoms in the cavity. Due to strong resonant interaction between two
atoms and optical cavity mode the absorbed input photon can be efficiently transferred from the
atoms to the resonant cavity mode. We show that the quantum transfer is highly sensitive to
the external excitation phases of atoms and it leads to quantum interference effects on the cavity
mode excitation. Besides, the quantum transfer is also influenced by the dipole-dipole interaction
dependent to the atomic distance. In this system the atomic positions also determine the coupling
constants between atoms and cavity mode which causes additional interference effects to the photon
exchange between atoms and cavity. Based on the characteristics of excitation phase we find that
it is a feasible scheme to generate long-lived dark state and it could be useful for storage and
manipulation of single photon fields by controlling the excitation phase.
\newline
~~~~~\newline
PACS number(s): 42.50.Pq, 42.50.Nn, 37.30.+i
\end{abstract}

\maketitle


\section{INTRODUCTION}
Recently, a lot of works have been carried out to investigate the phase effects induced by one single photon with many atoms \cite{Scully 06, Scully 07, Svidzinsky 08, Svidzinsky 08a, Scully 09s, Scully 09, Rohlsberger 10, Kessler 10, Svidzinsky 10, Svidzinsky 12,Slepyan 13, Miroshnychenko 13, Feng 14, Anatoly 15, Liao 15, kessel 93, Ohlsson 93, Moiseev 01, Riedmatten 08}. In these works the single photon not only supply energy to the system but also act as information (phase) carrier. A position-dependent excitation phase, dependent on the positions of excited atoms, is brought when atom in the ensemble is excited by one single photon. And such position-dependent excitation phase leads to a large number of quantum phenomena, such as directed spontaneous emission \cite{Scully 06, Scully 09s}, dynamical evolution of correlated spontaneous emission \cite{Svidzinsky 08}, collective lamb shift \cite{Scully 09}, quantum interference and quantum storage in photon echo media \cite{kessel 93, Ohlsson 93, Moiseev 01, Riedmatten 08}, etc. In order to describe this position-dependent excitation phase, timed Dicke state which contains the information of the excitation phase was introduced by Scully and his colleagues \cite{Scully 06, Scully 07}, through this state phase effects caused by one single photon can be explained and this kind of excitation phase can also be further understood.

It is well known that one of the factors effecting the coupling strength between light and matter is the light intensity. The coupling strength may be very weak when a single photon field interacts with matter, because the single photon has little intensity. However, by means of the cavity the coupling strength can be greatly enhanced even between one single photon and matter \cite{Reithmaier 04, William 06, Haroche 06}. The simplest model of the cavity QED is one atom cavity system, which is a well-known quantum system and is intensively used in quantum optics and quantum information science \cite{Goy 83, Maunz 04, Hetet 11, Haroche 06}. While further elaboration of single photon technologies push for the new quantum systems. Here two atoms in cavity promises to be very useful tool for quantum control of the single photon field, because the two-atom cavity system can realize more rich physical processes due to larger controllable Hilbert spaces. It is worth noting that the coherent control of two-atom cavity system has been already demonstrated in recent experiments \cite{Casabone 15, Reimann 15}.

In this paper, we investigate the quantum transport effects caused by excitation phase of one single photon in a two-atom cavity system. In this system two atoms located at different positions in an open cavity and they are excited by one external single photon with equal probability but with different excitation phases for their different positions. Due to the basic characteristic of cavity QED, energy exchange is existed between the cavity mode and atoms, the excitation phases brought by photon will be transferred between atoms and cavity mode with the photon exchange.

We show that the cavity mode could get a photon with different excitation phases which can lead to quantum interference effect. We have also demonstrated that the coherent photon transfer is highly effected by the different coupling constants due to the different atomic positions in the cavity, because in such situation new coupling relationships are introduced for coherent exchange of photon between the atoms and cavity mode. We analyze the dynamical and decoherent behaviors of this system with considering the interatomic dipole-dipole interaction and interaction with the free space light modes through the open cavity, which can greatly influence the evolution of two-atom cavity system. All of those properties are summarized by analyzing the dynamical behaviors of the dark state.

This paper is organized as follows. In Sec. II we present a scheme for preparation of the initial state in two-atom cavity system and analyze the states of the system under the conditions when two atoms have equal and unequal coupling constants with cavity mode. In Sec. III we give the dynamical behaviors of the system by using master-equation approach. Finally, we give our conclusion in section IV.

\section{PHYSICAL MODEL AND STATES OF SYSTEM}

\renewcommand{\figurename}{FIG.}
\begin{figure}[htbp]
\includegraphics[width=1\columnwidth]{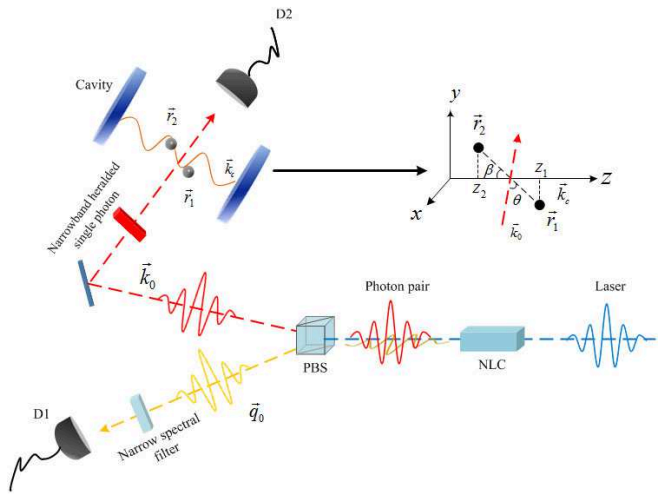}
\caption{(color online) Schematic representation of preparing a timed state in a cavity via a narrowband heralded single photon. The nonlinear crystal (NLC) down-converts one pump photon into a photon pair. A click in detector D1 indicates the generation of the photon pair, while no click in the detector D2 means the photon with the wave vector $\vec{k}_{0}$ is absorbed by two atoms. The preparation of a heralded single-photon field with narrowband spectrum and efficient absorption by a single atom in a free space scheme has been reliably experimentally observed \cite{Schuck 10, Piro 11, Piro 15} with the probability excitation close to one percent.}\label{fig:1}
\end{figure}

The principal scheme of the proposed experiment is depicted in Fig. \ref{fig:1}. Here two identical two-level atoms are located in an open cavity. One heralded single photon with wave vector $\vec{k}_{0}$ incidents from side of the cavity, which ensures the photon initially absorbed by the atoms instead of the cavity mode. If no photon detected (by the detector D2) at the emitting direction outside of the cavity we know that the single photon is absorbed by atoms, and the state of the system can be expressed as

\begin{equation}\label{1}
|\psi(0)\rangle=\frac{1}{\sqrt{2}}(|10\rangle|0\rangle e^{i\vec{k}_{0}\cdot\vec{r}_{1}}+|01\rangle|0\rangle e^{i\vec{k}_{0}\cdot\vec{r}_{2}}),
\end{equation}

\noindent where $|10\rangle$ ($|01\rangle$)indicates that the atoms with position $\vec{r}_{1}$ and $\vec{r}_{2}$ are in the excited (ground) and ground (excited) state, respectively, $|0\rangle$ indicates no photon in the cavity mode. The phase factor $e^{i\vec{k}_{0}\cdot\vec{r}_{i}}$ ($i=1,2$) shows that an atom with position vector $\vec{r}_{i}$ is excited by the photon with wave vector $\vec{k}_{0}$.

These two atoms are strongly coupled with the cavity mode, with free space (vacuum modes) outside the open cavity and interact with each other via the dipole-dipole interaction. For simplicity, in this section we consider only the interactions of atoms with the cavity mode, or the atoms with free space modes (in Appendix B) separately. We briefly describe the main properties of these interactions and present the quantum states which play important role in the evolution of two-atom cavity system. The interaction Hamiltonian of this atom-cavity system is

\begin{equation}\label{2}
 H_{I}=\hbar \sum_{i=1}^{2}(g_{i}\sigma_{i}^{+}a e^{i(\omega_{0}-\omega_{c})t}+g_{i}^{\ast}a^{\dag}\sigma_{i}e^{-i(\omega_{0}-\omega_{c})t}),
\end{equation}

\noindent where $\sigma_{i}^{+}$ ($\sigma_{i}$) is the raising (lowering) operator for the $i$th atom, $a^{\dag}$ and $a$ are field creation and annihilation operators of the cavity mode, $\omega_{0}$ and $\omega_{c}$ are the resonance frequency of atoms and cavity mode, respectively. $g_{i}$ is the coupling constant between $i$th atom and cavity mode and has the form of $g_{i}=g_{0}\cos (k_{c}z_{i})$ \cite{Zippilli 04, Vidal 04}, $k_{c}$ is the wave vector of the cavity, $z_{i}$ is the projection of the $r_{i}$ to the cavity axis and $z_{12}=r_{12}\cos\beta$, $\beta$ is the angle between $k_{c}$ and the atomic joining line $r_{12}$.

Due to the interaction between atoms and cavity, photon together with excitation phases will be transferred between atoms and cavity. But the dynamical behaviors of photon displays very different features for the cases of $g_{1}=g_{2}$ and $g_{1}\neq g_{2}$, which actually corresponds to different physical essences. In the following subsections we focus our discussion on these two conditions .

\subsection{Equal coupling constant $\mathbf{g_{1}=g_{2}}$ }

In order to investigate the dynamical evolution of the photon unitary time-evolution operator method \cite{Scully 97} is used, and the unitary time-evolution operation is given by

 \begin{equation}\label{3}
  U(t)=\exp(-i H_{I} t/\hbar).
 \end{equation}

\noindent For simplicity, we consider the simplest condition that the frequency of cavity mode $\omega_{c}$ is equal to the resonance frequency of the two-level atoms $\omega_{0}$, the analytic form of $U(t)$ under the condition of $g_{1}=g_{2}$ is given in Appendix A. The state of system at any time $t$ can be simply obtained by using the relation:

\begin{equation}\label{4}
 |\psi(t)\rangle=U(t)|\psi(0)\rangle.
\end{equation}

\noindent On substituting $|\psi(0)\rangle$ and $U(t)$ from Eqs. (\ref{1}) and (\ref{a1}) into Eq. (\ref{4}), we have

\begin{eqnarray}\label{5}
 |\psi(t)\rangle &=& c_{1}(t)\frac{1}{\sqrt{2}}(|00\rangle |1\rangle e^{i\vec{k}_{0}\cdot\vec{r}_{1}}+|00\rangle |1\rangle e^{i\vec{k}_{0}\cdot\vec{r}_{2}})+
 \notag \\&&c_{2}(t)\frac{1}{\sqrt{2}}(|10\rangle|0\rangle e^{i\vec{k}_{0}\cdot\vec{r}_{1}}+|01\rangle|0\rangle e^{i\vec{k}_{0}\cdot\vec{r}_{2}})+
 \notag \\&&c_{3}(t)\frac{1}{\sqrt{2}}(|10\rangle|0\rangle e^{i\vec{k}_{0}\cdot\vec{r}_{2}}+|01\rangle|0\rangle e^{i\vec{k}_{0}\cdot\vec{r}_{1}}),
\end{eqnarray}
\noindent where $c_{1}(t)=-i\frac{1}{\sqrt{2}}\sin\sqrt{2}gt$, $c_{2}(t)=\frac{1}{2}(\cos\sqrt{2}gt+1)$, $c_{3}(t)=\frac{1}{2}(\cos\sqrt{2}gt-1)$, and $|1\rangle$ contained in the first term indicates one photon in cavity mode.

Three terms are contained in Eq. (\ref{5}), which correspond to three different states. The state contained in the first term describes the excitation of the cavity mode and it includes two kinds of excitation phases which originate from absorbing photon with phases $e^{i\vec{k}_{0}\cdot\vec{r}_{1}}$ and $e^{i\vec{k}_{0}\cdot\vec{r}_{2}}$ from atoms located in the positions $\vec{r}_{1}$ and $\vec{r}_{2}$, respectively. From this term it can be seen that the cavity mode may absorb the photon with the same probability but different excitation phases. We calculate the probability of cavity mode to show the dynamical behavior of the photon:

\begin{equation}\label{6}
 P=\langle E^{(-)}E^{(+)}\rangle,
\end{equation}
\noindent where $E^{(-)}=\mathscr{E} a^{\dag}$ ($\mathscr{E}$ is the amplitude of the cavity mode) and $E^{(+)}$ is conjugate of $E^{(-)}$, substituting  Eq. (\ref{5}) into Eq. (\ref{6})

\begin{eqnarray}\label{7}
 P&=&\langle\psi(t)|E^{(-)}E^{(+)}|\psi(t)\rangle
 \notag \\&=& \mathscr{E}^{2}|c_{1}(t)|^{2}(1+\cos[\vec{k}_{0}\cdot(\vec{r}_{2}-\vec{r}_{1})]),
\end{eqnarray}

\noindent where the interference term exhibits a cosine modulation with relative phase $\varphi=\vec{k}_{0}\cdot(\vec{r}_{2}-\vec{r}_{1})$. Thus the change of the wave vector $k_{0}$ will provide a coherent control of the photon emission in the cavity mode.

The state contained in the second term is identical to the initial state of Eq. (\ref{1}) and this corresponds to a process that the photon is firstly transferred from an atom (e.g with position vector $\vec{r}_{i}$) to the cavity mode and then transferred back to the same atom ( still the atom with position vector $\vec{r}_{i}$). The state described by the third term has the similar expression with Eq. (\ref{1}) but with the excitation phases interchanged, which indicates that a photon is firstly transferred from an atom (e.g with position vector $\vec{r}_{i}$) to the cavity mode but transferred back to another atom (the atom with position vector $\vec{r}_{j}$ with $i\neq j$). Actually the last two states of Eq. (\ref{5}) can be separated into two parts, one is decoupled from the cavity mode (time-independent terms) and we use $|\psi\rangle_{a}$ to express it
\begin{eqnarray}\label{8}
 |\psi\rangle_{a} &=& \frac{1}{2\sqrt{2}}(|10\rangle|0\rangle -|01\rangle|0\rangle)(e^{i\vec{k}_{0}\cdot\vec{r}_{1}}-e^{i\vec{k}_{0}\cdot\vec{r}_{2}})
 \notag \\&=&\frac{1}{2}|D\rangle e^{i\vec{k}_{0}\cdot\vec{r}_{1}}(1-e^{i\varphi}).
\end{eqnarray}

\noindent where $|D\rangle=\frac{1}{\sqrt{2}}(|10\rangle|0\rangle -|01\rangle|0\rangle)$ is also the dark state of the free space (see Appendix B) and it is decoupled from the cavity mode under the condition of $g_{1}=g_{2}$ \cite{Nicolosi 04}. The remaining part is coupled with the cavity mode and we use $|\psi\rangle_{b}$ to express it

\begin{eqnarray}\label{9}
 |\psi\rangle_{b} &=& \frac{1}{2\sqrt{2}}\cos\sqrt{2}gt(|10\rangle|0\rangle +|01\rangle|0\rangle)(e^{i\vec{k}_{0}\cdot\vec{r}_{1}}+e^{i\vec{k}_{0}\cdot\vec{r}_{2}})
 \notag \\&=&\frac{1}{2}\cos\sqrt{2}gt|B\rangle e^{i\vec{k}_{0}\cdot\vec{r}_{1}}(1+e^{i\varphi}).
\end{eqnarray}

\noindent where $|B\rangle=\frac{1}{\sqrt{2}}(|10\rangle|0\rangle +|01\rangle|0\rangle)$ is the bright state of the free space (see Appendix B), which is coupled with the cavity mode and can exchange energy with the cavity mode under the condition of $g_{1}=g_{2}$.

In the basis of $|D\rangle$ and $|B\rangle$, the initial state (\ref{1}) can be decomposed into two components

\begin{equation}\label{10}
|\psi(0)\rangle=a|D\rangle+b|B\rangle,
\end{equation}

\noindent where $a=\frac{1}{2}e^{i\vec{k}_{0}\cdot\vec{r}_{1}}(1-e^{i\varphi})$, $b=\frac{1}{2}e^{i\vec{k}_{0}\cdot\vec{r}_{1}}(1+e^{i\varphi})$. Apparently the dark state $|D\rangle$ shown in Eq. (10) is decoupled with both cavity mode and vacuum under the condition of $g_{1}=g_{2}$, and therefore photon assumes to be conserved in this state for a very long time. In this atom-cavity system only the bright state $|B\rangle$ exchanges photon and excitation phases with the cavity mode and finally leads to the quantum interference effects.

\subsection{Unequal coupling constant $\mathbf{g_{1} \neq g_{2}}$}

For two atoms with different positions, they may posses different excited phases during the excited process as shown in Eq. (\ref{1}), while for a standing wave cavity the different atomic positions may also lead to the different coupling strengths, i.e., $g_{1}\neq g_{2}$. The simple analytic form for $U(t)$ in the condition of $g_{1}\neq g_{2}$ can not be obtained, but different physical essences can be revealed by comparing the dynamical behaviors of the dark state at the conditions of $g_{1}= g_{2}$ and $g_{1}\neq g_{2}$. By using the interaction Hamiltonian of Eq. (\ref{2}) the dark state of the cavity can be obtained according to the following definition

\begin{equation}\label{11}
\hbar(g_{1}a^{\dag}\sigma_{1}+g_{2}a^{\dag}\sigma_{2})|D\rangle_{c}=0,
\end{equation}

\noindent where we still only consider the resonance condition $\omega_{0}=\omega_{c}$. Here $|D\rangle_{c}$ is used to present the dark state of the cavity mode and it can be expressed as

\begin{equation}\label{12}
|D\rangle_{c}=\frac{1}{\sqrt{g_{1}^{2}+g_{2}^{2}}}(g_{2}|10\rangle|0\rangle-g_{1}|01\rangle|0\rangle).
\end{equation}

\noindent
 Unlike the dark sate $|D\rangle$ shown in Eq. (\ref{8}) the dark state of the cavity mode of Eq. (\ref{12}) is related to the coupling constants $g_{i}~(i=1,2)$ when $g_{1} \neq g_{2}$. The bright state of the cavity mode can be obtained according to $_{c}\langle D|B\rangle_{c}=0$

\begin{equation}\label{13}
|B\rangle_{c}=\frac{1}{\sqrt{g_{1}^{2}+g_{2}^{2}}}(g_{1}|10\rangle|0\rangle+g_{2}|01\rangle|0\rangle).
\end{equation}

\noindent We can infer that when $g_{1}=g_{2}$ the dark and bright states of cavity mode $|D\rangle_{c}$ and $|B\rangle_{c}$ are the same with the ones of free space $|D\rangle$ and $|B\rangle$, because in the free space each atom also has the equal coupling strength with the vacuum modes. However, when $g_{1}\neq g_{2}$ they are different from the ones of free space, because in this situation two atoms have unequal coupling strengths with the cavity mode but equal coupling strength with the vacuum mode. So when $g_{1}=g_{2}$ the dark and bright states of free space ($|D\rangle$ and $|B\rangle$) and cavity mode ($|D\rangle_{c}$ and $|B\rangle_{c}$) have the same expression and do not need to distinguish, while when $g_{1}\neq g_{2}$  the dark and bright states of free space and cavity mode are different and should be distinguished.

Due to the difference of $g_{1}$ and $g_{2}$ Eqs. (\ref{12}) and (\ref{13}) can also be written as

\begin{equation}\label{14a}
|D\rangle_{c}=\frac{1}{\sqrt{g_{1}^{2}+g_{2}^{2}}}[\frac{(g_{2}-g_{1})}{\sqrt{2}}|B\rangle+\frac{(g_{1}+g_{2})}{\sqrt{2}}|D\rangle],
\end{equation}

\begin{equation}\label{14b}
|B\rangle_{c}=\frac{1}{\sqrt{g_{1}^{2}+g_{2}^{2}}}[\frac{(g_{1}+g_{2})}{\sqrt{2}}|B\rangle+\frac{(g_{1}-g_{2})}{\sqrt{2}}|D\rangle],
\end{equation}

\noindent which are superposition states of the dark state $|D\rangle$ and bright state $|B\rangle$. From Eq. (\ref{14b}) we know that the dark state $|D\rangle$ is no longer decoupled with cavity mode when $g_{1}\neq g_{2}$, therefore in this condition the dark state $|D\rangle$ also exchanges energy with the cavity mode. What's more, from Eq. (\ref{14b}) we can conclude that the states $|D\rangle$ and $|B\rangle$ could get coupled via the cavity mode. All of which can greatly effect the dynamical behavior of the cavity mode and atoms.

\section{EVOLUTION OF TWO-ATOM CAVITY SYSTEM}

\subsection{Photon transport in two-atom cavity}

In Section II we discussed the two-atom cavity system with considering the effects of cavity and free space separately. However, for an open cavity system the atoms interact with the cavity mode and free space mode simultaneously. We analyze these effects in the framework of master equation approach.

In this system the distance between two atoms is closer than the radiation wavelength where the dipole-dipole interaction of two atoms \cite {Agrwal 74} will play a significant role in the analyzed processes. The master equation of this atom-cavity system can be expressed as

\begin{eqnarray}\label{15}
\frac{d}{dt}\rho&=&-i\omega_{0}\sum_{i=1}^{2}[\sigma_{i}^{z},\rho]-i\omega_{c}[a^{\dag}a,\rho]-i\sum_{i\neq j}\Omega_{ij}[\sigma_{i}^{\dag}\sigma_{j},\rho] \notag \\
&&-i \sum_{i=1}^{2}[g_{i}a\sigma_{i}^{\dag}+H.c,\rho]-\sum_{i,j=1}^{2}\gamma_{ij}(\rho\sigma_{i}^{\dag}\sigma_{j}+\sigma_{i}^{\dag}\sigma_{j}\rho\notag \\&&-2\sigma_{j}\rho\sigma_{i}^{\dag})-\kappa(\rho a^{\dag}a+ a^{\dag}a\rho-2a\rho a^{\dag}),
\end{eqnarray}
\noindent where $\sigma_{i}^{z}=(\sigma_{i}^{\dag}\sigma_{i}-\sigma_{i}\sigma_{i}^{\dag})/2$ is the energy operator for $i$th atoms, $\Omega_{ij}$ and $\gamma_{ij}$ are the dipole-dipole interaction term and the cooperative decay rate, respectively, $\kappa$ describes the cavity losses.

The expression of dipole-dipole interaction and the cooperative decay rate are \cite{Agrwal 74, Rudolph 95}

\begin{eqnarray}\label{16}
\Omega_{ij}&=&\frac{3}{2}\gamma\{-(1-\cos^{2}\alpha)\frac{\cos(k r_{ij})}{k r_{ij}}\notag \\
&&+(1-3\cos^{2}\alpha)[\frac{\sin(k r_{ij})}{(k r_{ij})^{2}}+\frac{\cos(k r_{ij})}{(k r_{ij})^{2}}]\},
\end{eqnarray}
\begin{eqnarray}\label{17}
\gamma_{ij}&=&\frac{3}{2}\gamma\{(1-\cos^{2}\alpha)\frac{\sin(k r_{ij})}{k r_{ij}}\notag \\
&&+(1-3\cos^{2}\alpha)[\frac{\cos(k r_{ij})}{(k r_{ij})^{2}}-\frac{\sin(k r_{ij})}{(k r_{ij})^{2}}]\},
\end{eqnarray}

\noindent where $2\gamma=2\gamma_{11}=2\gamma_{22}=4|\vec{d}_{eg}|^{2}\omega^{3}/3\hbar c^{3}$ is the spontaneous decay rate of the individual atom, $\vec{d}_{eg}$ is the dipole moment, $r_{ij}=|\vec{r}_{i}-\vec{r}_{j}|$ is the distance between the two atoms, $\alpha$ is the angle between the dipole moment $\vec{d}_{eg}$ and the vector $r_{ij}$, $k=\omega/ c$.

Similar to \cite{Sumanta 08, Eyob 11}, in our system the excited atoms bring the position dependent excitation phase excited by a single photon. But general form of the master equation Eq. (\ref{15}) does not contain information about the initial state of two atoms. We can easily account for the initial state by using the basis atomic states that contain the initial phases of the excited atoms.

\begin{eqnarray}\label{18}
|g\rangle=|00\rangle|0\rangle,|e_{1}\rangle=|10\rangle|0\rangle e^{i\vec{k}_{0}\cdot\vec{r}_{1}},\nonumber \\
|e_{2}\rangle=|01\rangle|0\rangle e^{i\vec{k}_{0}\cdot\vec{r}_{2}},|c\rangle=|00\rangle|1\rangle,
\end{eqnarray}

\noindent where $|g\rangle$ is the ground state of the system, $|e_{i}\rangle$ ($i=1,2$) means that atom with position vector $\vec{r}_{i}$ is excited and brings an position-dependent excitation phase $e^{i\vec{k}_{0}\cdot\vec{r}_{i}}$, $|c\rangle$ indicates the excitation state of cavity mode. By means of these basis states the dynamical behaviors of the two-atom cavity system can be investigated through density matrix equations given in Appendix C. Especially, we are interested in the dynamical behaviors of cavity mode and atoms when the excitation phase is considered.

\begin{figure}[htbp]
\includegraphics[width=1\columnwidth]{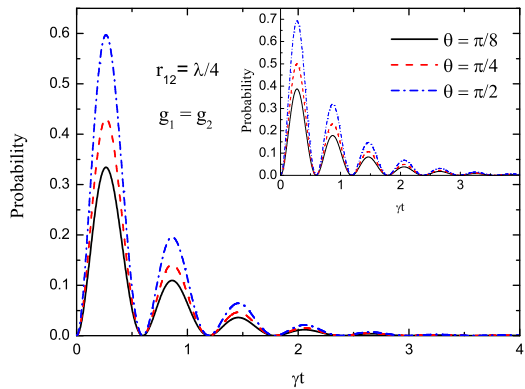}
\caption{(color online) The time-evolution of probability of finding the input photon in the cavity mode under different angle $\theta$ (the angle between wave vector $\vec{k}_{0}$ and $\vec{r}_{1}-\vec{r}_{2}$). The inset figure does not consider the dipole-dipole interaction, i.e. $r_{ij}=0$, $\Omega_{ij}=0$. Other parameters: $r_{12}=\lambda/4$, $g_{0}=5\gamma$, $g_{1}=g_{2}=g_{0}\cos(k_{0}\frac{r_{12}}{2}\cos\beta)$, $\beta$ is the angle between $k_{c}$ and the atomic joining line $r_{12}$ and here $\beta=\pi/8$,  $\alpha=\pi/2$, $\Delta=0$ and $\kappa=0.3\gamma$.}\label{fig:2}
\end{figure}

In Fig. \ref{fig:2} we give the probability of finding the input photon in the cavity mode under the condition of $g_{1}=g_{2}$ and $r_{12}=\lambda/4$. As shown in Fig. \ref{fig:2} the probability of cavity mode show oscillation and has the maximum value in each period with $\theta=\pi/2$, and the amplitude decreases with $\theta$ decreasing. These phenomena match the result given in Eq. (\ref{7}), when $\theta=\pi/2$ the interference term $\cos[\vec{k}_{0}\cdot(\vec{r}_{1}-\vec{r}_{2})]$ has value of 1, which corresponds to the completely constructive interference. When $\theta=\pi/4$ and $\theta=\pi/8$ the values of the interference term are smaller than 1, which corresponds to the incompletely constructive interference, thus the amplitudes of cavity mode in these cases are smaller than the case of $\theta=\pi/2$. The only changed parameter in Fig. \ref{fig:2} is the angle $\theta$ and the different angle $\theta$ actually introduce different excitation phases, which correspond to different degrees of quantum interference on the cavity mode. Such quantum interference can also be explained by Eq. (\ref{10}), with $g_{1}=g_{2}$ only bright state $|B\rangle$ couples to the cavity mode and the probability amplitude of this state is related to the angle $\theta$.

\begin{figure}[htbp]
\includegraphics[width=1\columnwidth]{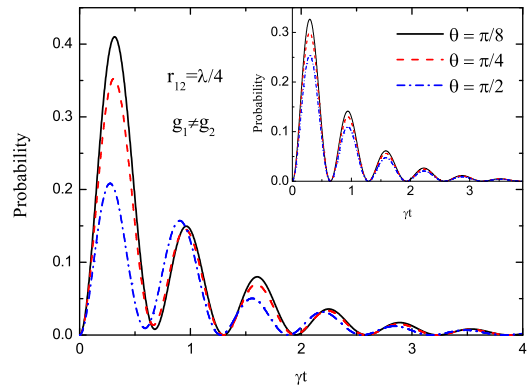}
\caption{(color online) The time-evolution of probability of finding the input photon in the cavity mode under different angle $\theta$. The inset figure does not consider the dipole-dipole interaction, i.e. $r_{ij}=0$, $\Omega_{ij}=0$. Other parameters: $r_{12}=\lambda/4$, $g_{0}=5\gamma$, $g_{1}=g_{0}\cos[k_{0}(\frac{\lambda}{6}-\frac{r_{12}}{2})\cos\beta]$, $g_{2}=g_{0}\cos[k_{0}(\frac{\lambda}{6}+\frac{r_{12}}{2})\cos\beta]$, $\beta=\pi/8$, $\alpha=\pi/2$, $\Delta=0$ and $\kappa=0.3\gamma$. }\label{fig:3}
\end{figure}

In Fig. \ref{fig:3} we give probability of finding the input photon in the cavity mode under the condition of $g_{1}\neq g_{2}$. The probability of cavity mode still show oscillation forms, but when $\theta=\pi/2$ the cavity mode has the minimum value in each period, which is very different from Fig. \ref{fig:2}. Because when $g_{1}\neq g_{2}$ the amplitudes of the cavity mode not only dependent to the angle $\theta$ but also sensitive to the coupling constants between atoms and cavity mode. When $g_{1}\neq g_{2}$ the state of the cavity mode would be

\begin{equation}\label{19}
|c\rangle=\frac{\alpha(g_{1},g_{2},t)}{\sqrt{g_{1}^{2}+g_{2}^{2}}}(g_{1}|00\rangle|1\rangle e^{i\vec{k}_{0}\cdot\vec{r}_{1}}+g_{2}|00\rangle|1\rangle e^{i\vec{k}_{0}\cdot\vec{r}_{2}}),
\end{equation}

\noindent where $\alpha(g_{1},g_{2},t)$ is a function of $g_{1}$, $g_{2}$ and $t$. According to Eq. (\ref{6}) the probability of cavity mode absorbing a photon is

\begin{equation}\label{20}
 P= \mathscr{E}^{2}\frac{\alpha^{2}}{g_{1}^{2}+g_{2}^{2}}(g_{1}^{2}+g_{2}^{2}+g_{1}g_{2}\cos[\vec{k}_{0}\cdot(\vec{r}_{2}-\vec{r}_{1})]),
\end{equation}

\noindent and we can infer that the quantum interference displays very differently under the conditions of $g_{1}g_{2}>0$ and $g_{1}g_{2}<0$. In Fig. \ref{fig:2} with $g_{1}g_{2}>0$ constructive interference is displayed with the increase of the angle $\theta$, while in Fig. \ref{fig:3} with $g_{1}g_{2}<0$ it presents destructive interference with the increase of the angle $\theta$. On the other hand we notice that the lifetime of the cavity mode in Fig. \ref{fig:3} is longer than that in Fig. \ref{fig:2} though the cavity mode has a smaller amplitudes in Fig. \ref{fig:3}. The reason is that in Fig. \ref{fig:2} with $g_{1}=g_{2}$ only bright state $|B\rangle$ couples to the cavity mode, while in Fig. \ref{fig:3} with $g_{1}\neq g_{2}$ both the dark state $|D\rangle$ and bright state $|B\rangle$ couple to the cavity mode. Therefore when $g_{1}\neq g_{2}$ the energy exchange is also existed between dark state $|D\rangle$ and cavity mode and it is known that the lifetime of the dark state $|D\rangle$ is longer than the bright state $|B\rangle$.

Finally, when analyzing the properties of photon exchange in the two-atom cavity system, we should notice a visible influence of the dipole-dipole interaction to the amplitude and period of the photon oscillations. In particular, in Fig. \ref{fig:2} the dipole-dipole interaction decreases the oscillation amplitude. The situation is complicated in Fig. \ref{fig:3}, where the dipole-dipole interaction decreases the oscillation amplitude when $\theta=\pi/2$ but increases the oscillation amplitude when $\theta=\pi/8$ and $\theta=\pi/4$. Besides, in Fig. \ref{fig:3} the oscillation period is also changed by the dipole-dipole interaction, for example, when $\theta=\pi/2$ the cavity mode has faster oscillation in the first period and slower oscillation in the second period then similar situation repeats in the following periods. While when $\theta=\pi/4$ and $\theta=\pi/8$ oscillation velocity in each period is opposite to the one when $\theta=\pi/2$.

\begin{figure}[htbp]
\includegraphics[width=1\columnwidth]{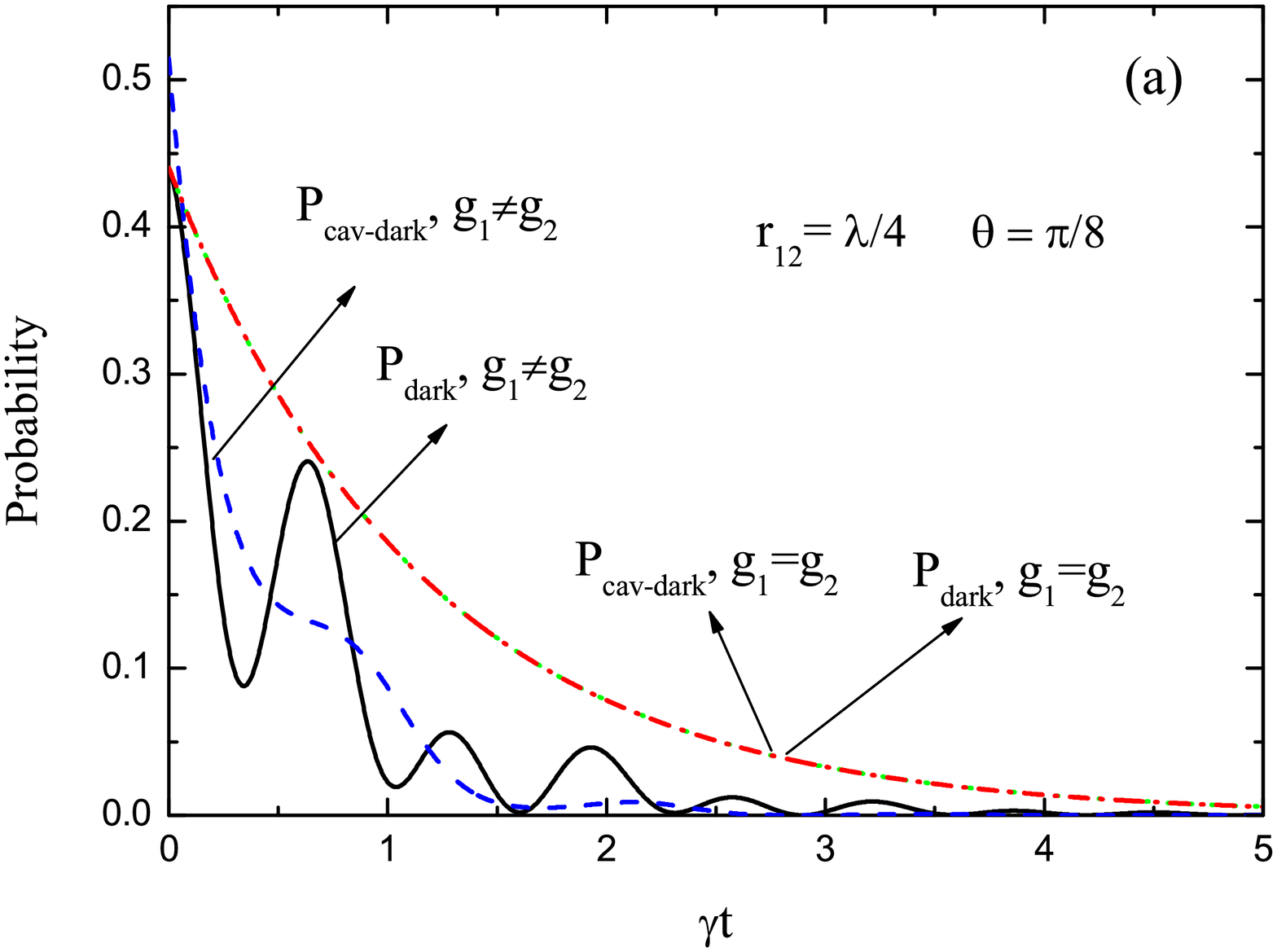}
\includegraphics[width=1\columnwidth]{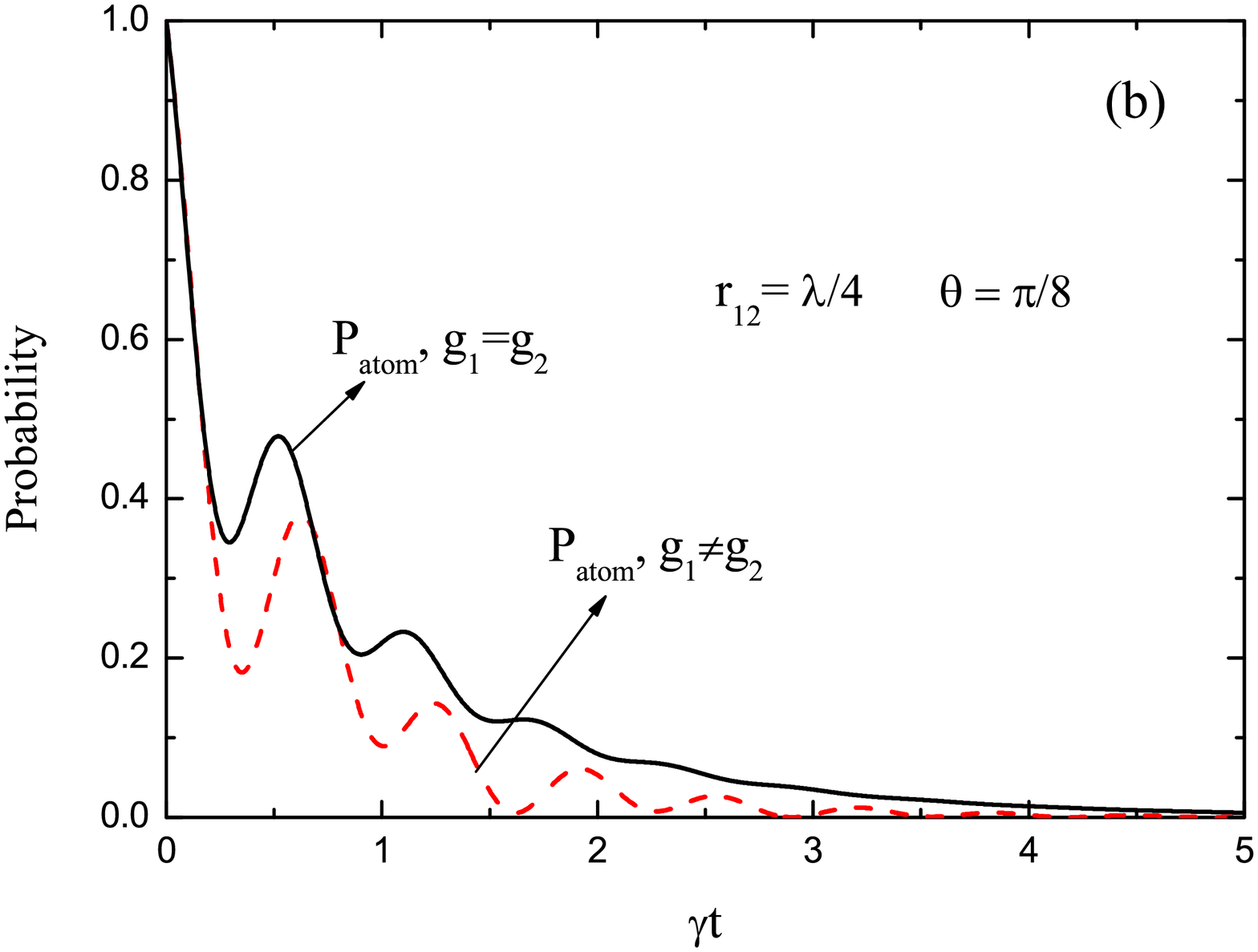}
\caption{(color online) The probability of dark states and atom under the conditions of $g_{1}=g_{2}$ ($g_{1}=g_{2}=g_{0}\cos(k_{0}\frac{r_{12}}{2}\cos\beta)$) and $g_{1}\neq g_{2}$ ( $g_{1}=g_{0}\cos[k_{0}(\frac{\lambda}{6}-\frac{r_{12}}{2})\cos\beta]$, $g_{2}=g_{0}\cos[k_{0}(\frac{\lambda}{6}+\frac{r_{12}}{2})\cos\beta]$). Other parameters: $\theta=\pi/8$, $r_{12}=\lambda/4$, $g_{0}=5\gamma$,  $\beta=\pi/8$, $\alpha=\pi/2$, $\Delta=0$ and $\kappa=0.3\gamma$. In Fig. (\ref{fig:4}) $P_{cav-dark}$ and $P_{dark}$ are the population of $|D\rangle_{c}$ and $|D\rangle$, respectively. $P_{atom}$ is the population of atoms.}\label{fig:4}
\end{figure}

In order to further illustrate the difference between the conditions of $g_{1}=g_{2}$ and $g_{1}\neq g_{2}$, time-evolutions of dark states and atoms are given in Fig. \ref{fig:4}(a) and Fig. \ref{fig:4}(b). In Fig. \ref{fig:4}(a) when $g_{1}=g_{2}$, dark states $|D\rangle$ and $|D\rangle_{c}$ decay exponentially with time and the two curves are overlapping as we analyzed in Section II. However, with $g_{1}\neq g_{2}$ the dark states $|D\rangle$ and $|D\rangle_{c}$ are non-overlapping and decay faster than the case of $g_{1}=g_{2}$. Besides, when $g_{1}\neq g_{2}$ the curves of $|D\rangle$ and $|D\rangle_{c}$ show oscillation forms.

These difference can be explained by means of Eq. (\ref{14b}), usually the dark state $|D\rangle$ decouples to the cavity mode but when $g_{1}\neq g_{2}$ it couples to the cavity mode, so energy exchange is existed between the dark state $|D\rangle$ and the cavity mode when $g_{1}\neq g_{2}$ and oscillation is appeared. From Eq. (\ref{14b}) we also notice that with $g_{1}\neq g_{2}$ the dark state $|D\rangle$ also get coupled with the bright state $|B\rangle$ via the cavity mode, therefore the dark state $|D\rangle$ can decay to the vacuum through both the cavity losses and the bright state which leads to its faster decay. Similar situation happens to the cavity mode dark state $|D\rangle_{c}$, when $g_{1}\neq g_{2}$ the $|D\rangle_{c}$ is the dark state of the cavity mode but not the free space. In this situation the dark state $|D\rangle_{c}$ couples to the bright state $|B\rangle$ and decays to the vacuum through the bright state $|B\rangle$, which leads to its faster decay in comparison with the case of $g_{1}= g_{2}$. The oscillation shown on the $|D\rangle_{c}$ is caused by dipole-dipole interaction but not the energy exchange between the cavity mode, because $|D\rangle_{c}$ is decoupled with the cavity mode.

In Fig. \ref{fig:4}(b) the probability of photon in atomic excitation state shows oscillation forms under the conditions of $g_{1}=g_{2}$ and $g_{1}\neq g_{2}$, but when $g_{1}\neq g_{2}$ the atoms decay faster than at the condition $g_{1}=g_{2}$. The population of atoms can be expressed as

\begin{eqnarray}\label{21}
 P_{atom}&=&P_{dark}+P_{bright}\notag
 \\&=& P_{cav-dark}+P_{cav-bright},
\end{eqnarray}

\noindent where $P_{dark}$ and $P_{bright}$ are the population of dark state $|D\rangle$ and bright state $|B\rangle$. $P_{cav-dark}$ and $P_{cav-bright}$ are the population of dark state $|D\rangle_{c}$ and bright state $|B\rangle_{c}$. When $g_{1}=g_{2}$ dark state $|D\rangle$ and bright state $|B\rangle$ are decoupled and these two components decay independently, the bright state decays faster than the dark state and finally only the dark state left. However, when $g_{1}\neq g_{2}$ dark state $|D\rangle$ and bright state $|B\rangle$ get coupled via the cavity mode. The decay process of the dark state $|D\rangle$ is affected by the bright state $|B\rangle$ and the decay rate of atoms is larger compare with the condition of $g_{1}=g_{2}$, which leads to the faster decay of atoms. In Fig. \ref{fig:4}(b) we also notice that the oscillation under the condition of $g_{1}\neq g_{2}$ last longer than the condition of $g_{1}=g_{2}$ and similar phenomenon also happens in the cavity mode of Fig. \ref{fig:3}. Here this phenomenon is also caused by the energy exchange between dark state $|D\rangle$ and cavity mode.

\subsection{Generation of two-atom dark state}

\begin{figure}[htbp]
\includegraphics[width=1\columnwidth]{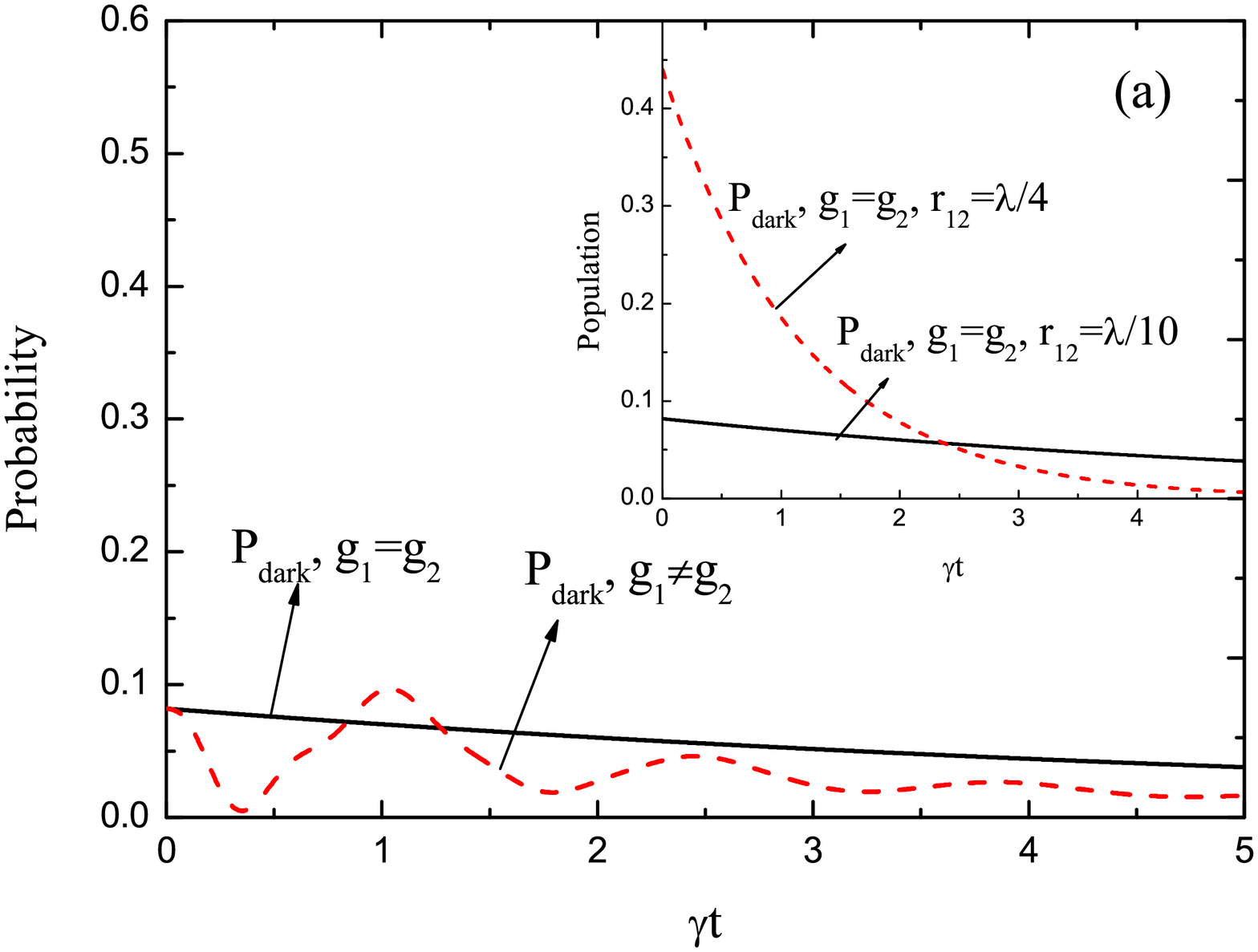}
\includegraphics[width=1\columnwidth]{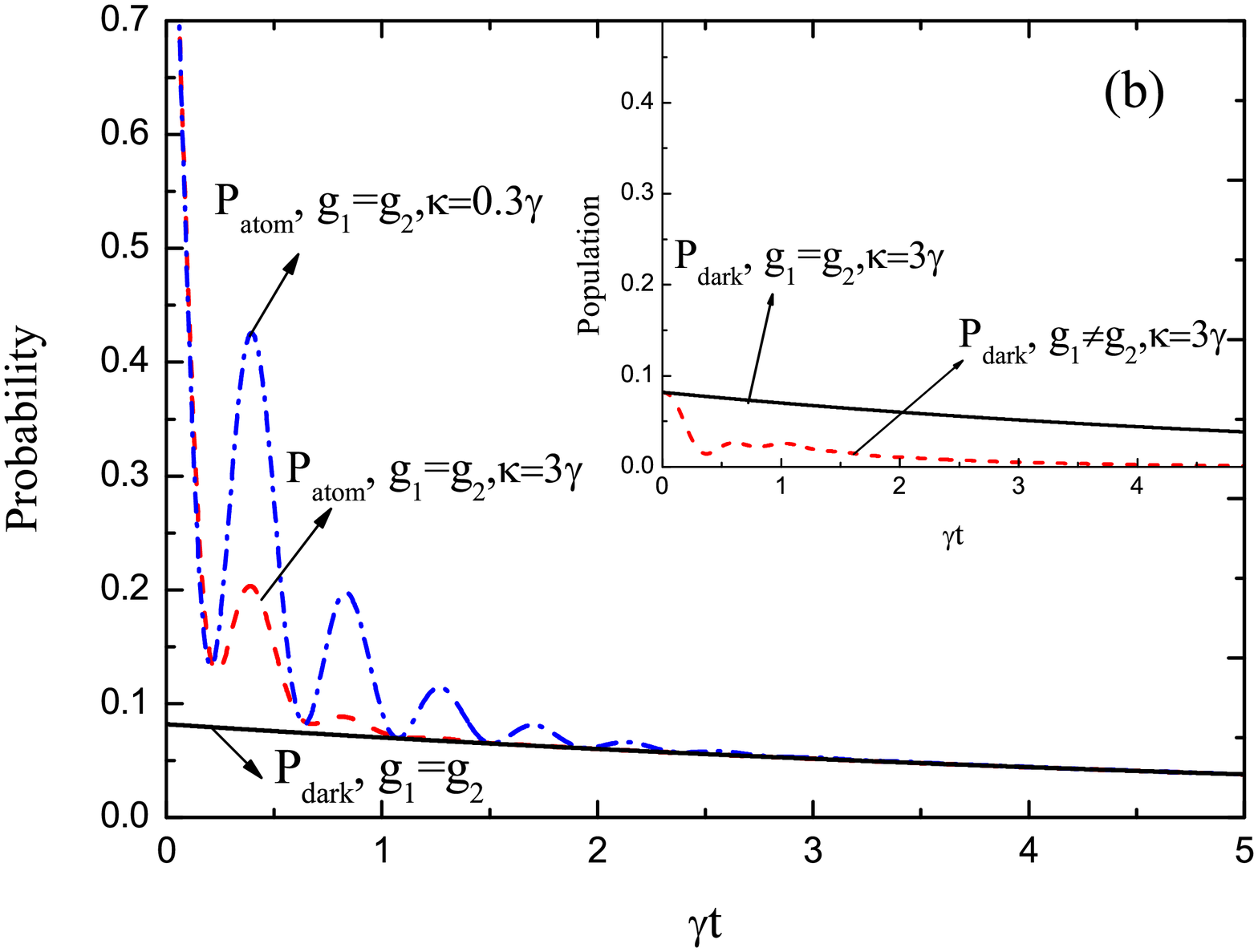}
\caption{(color online) The probability of dark states and atom under the conditions of $g_{1}=g_{2}$ ($g_{1}=g_{2}=g_{0}\cos(k_{0}\frac{r_{12}}{2}\cos\beta)$) and $g_{1}\neq g_{2}$ ( $g_{1}=g_{0}\cos[k_{0}(\frac{\lambda}{6}-\frac{r_{12}}{2})\cos\beta]$, $g_{2}=g_{0}\cos[k_{0}(\frac{\lambda}{6}+\frac{r_{12}}{2})\cos\beta]$). Other parameters: $r_{12}=\lambda/10$, $\theta=\pi/8$, $g_{0}=5\gamma$, $\beta=\pi/8$, $\alpha=\pi/2$, $\Delta=0$ and $\kappa=0.3\gamma$. The inset of Fig. \ref{fig:5}(a) show the dynamical behavior of dark state $|D\rangle$ under the atomic distances of $r_{12}=\lambda/10$ and $r_{12}=\lambda/4$. The inset of Fig. \ref{fig:5}(b) show the dynamical behavior of dark state $|D\rangle$ affected by different cavity decay rates $\kappa=0.3\gamma$ and $\kappa=3\gamma$.}\label{fig:5}
\end{figure}

 As was shown above the two atom cavity system excited by a heralded single photon can demonstrate a variety of quantum dynamics scenario. This is of particular interest to consider the generation of long-lived dark state in these processes. In Fig. \ref{fig:5} we analyze the factors which could affect the generation of dark state.

In Fig. \ref{fig:5} we present the population probability of the dark state under the distance of $r_{12}=\lambda/10$ and $\theta=\pi/8$. The reason for choosing such small atomic distance is that the small atomic distance can increase the decay rate of bright state and decrease the decay rate of the dark state. The population probability of the dark state under the condition of $r_{12}=\lambda/4$ and $r_{12}=\lambda/10$ is compared in inset of Fig. \ref{fig:5}(a). The initial value of dark state is larger when $r_{12}=\lambda/4$ but it decays very quickly and almost 0 at $\gamma t=5$. This means that under a given excitation phase the generation of the dark state is strongly influenced by the dipole-dipole interaction related to the different atomic distances. In Fig. \ref{fig:5}(a) we give the dynamical behavior of dark state under the condition of $g_{1}=g_{2}$ and $g_{1}\neq g_{2}$. As shown in Fig. \ref{fig:5}(a) the dynamical behavior of dark state is a straight line with small decay rate when $g_{1}=g_{2}$. While when $g_{1}\neq g_{2}$ oscillation form is displayed on the line of dark state, which violate the stability of dark state should be.

In Fig. \ref{fig:5}(b) we present the effect brought by cavity losses. As shown in Fig. \ref{fig:5}(b) the population of atoms $P_{atom}$ oscillates at the beginning due to the periodic energy exchange between atoms and cavity mode then it tends to the line of $P_{dark}$. That's because the initial state of Eq. (\ref{1}) contains components of dark state $|D\rangle$ and bright state $|B\rangle$, the bright state decay faster than the dark state and finally only the dark state left in this system. The cavity decay rate determines the time to obtain a dark state and it needs shorter time to get a pure dark state in a bad cavity. As shown in Fig. \ref{fig:5}(b) after $\gamma t\geq 3$ almost a pure of dark state can be obtained with $\kappa=0.3\gamma$, while this time can be greatly decreased when $\kappa=3\gamma$. This happens because the large cavity losses lead to a faster decay for the bright state but has no effect on the dark state. However, the cavity losses can effect the dynamical behavior of dark state when $g_{1}\neq g_{2}$, the inset of Fig. \ref{fig:5}(b) shows that with $g_{1}\neq g_{2}$ the dark state will decay rapidly and we can not obtain a dark state under this situation.

\section{CONCLUSION}

Two atom system is a promising system that attracts an increasing wave of attentions \cite{Casabone 15, Reimann 15, Zheng 00, Ficek 02, Wilk 10, Berman 15, Milonni 15}. In this paper, two atoms are located in an optical cavity. We study the phase effects in a coherent transport of a single photon excitation in this system that is induced by external heralded single photon. We show that this process seems to be interesting for efficient coherent control and storage of single photons by manipulating the initial excitation phase of two atoms.

We first analyze the role of the prepared initial quantum state and we find that the excitation atomic phases can lead to strong quantum interference effects on the cavity mode excitation. We further show that such quantum interference of the intracavity photon is not only effected by the excitation phase but also controlled by the coupling constants between atoms and cavity. When the coupling constants are different for two atoms, interesting physical phenomena are displayed in the dynamical behavior of cavity mode and atoms. All of these phenomena are caused by new coupling relationships between atoms and cavity mode. Especially, when the coupling constants between atoms and cavity mode are different, the dark state of free space couples to the cavity mode and energy exchange is existed between such dark state and cavity mode. In addition, due to the existence of excitation phase the initial state can be decomposed into dark and bright states, which can present a feasible scheme to generate dark state through this two-atom cavity system using one narrowband heralded single photon.

Also we have found that the dipole-dipole interaction plays an important role in the photon transfer in the two-atom cavity system. In particular this interaction can influence the photon exchange and for small interaction distance it provides stronger generation of the long-lived two atomic dark states. However, clearer understanding the role of dipole-dipole interaction to the properties of two-atom cavity system requires further studies that will be a subject of future investigation.

\section{ACKNOWLEDGMENT}
We acknowledge support on the National Natural Science Foundation of China (Grant NO.11174109).

Sergey A. Moiseev thanks support of High-end Foreign Experts from the State Administration of Foreign Experts Affairs of China (Grant No, GDW 20152200077) and Russian Foundation for Basic Research (Grant No.14-02-00903/14).

\renewcommand{\theequation}{A.\arabic{equation}}
\setcounter{equation}{0}  
\section*{APPENDIX A: UNITARY TIME-EVOLUTION OPERATION}  

Substituting Eq. (\ref{3}) into Eq. (\ref{2}) and after a long and tedious calculation Eq. (\ref{2}) has the expression as

 \begin{eqnarray}\label{a1}
 U(t)&=&\frac{1}{2}[\cos(gtC)\sum_{i=1}^{2}|e_{i}\rangle\langle e_{i}|+\cos(gt\sqrt{2a^{\dag}a})\sum_{i=1}^{2}|g_{i}\rangle\langle g_{i}|
 \notag \\&&+\cos(gtC)\sum_{i\neq j}\sigma_{i}^{+}\sigma_{j}+\cos(gt\sqrt{2a^{\dag}a})\sum_{i\neq j}\sigma_{i}^{+}\sigma_{j}]
 \notag \\&&-i\frac{1}{C}\sin(gtC)a\sum_{i=1}^{2}\sigma_{i}^{+}-ia^{\dag}\frac{1}{C}\sin(gtC)\sum_{i=1}^{2}\sigma_{i}
 \notag \\&&-\sum_{i\neq j}\sigma_{i}^{+}\sigma_{j}.
\end{eqnarray}
\noindent where $C=\sqrt{2+2a^{\dag}a}$.

\renewcommand{\theequation}{B.\arabic{equation}}
\setcounter{equation}{0}  
\section*{APPENDIX B: THE DARK STATE OF THE FREE SPACE}

In Fig. \ref{fig:1} atoms can interact with the free space mode through the open cavity. The interaction Hamiltonian which considers the dipole-dipole interaction is

\begin{equation}\label{b1}
 H_{I}=\hbar\Omega_{12}(\sigma_{1}^{+}\sigma_{2}+\sigma_{2}^{+}\sigma_{1}).
\end{equation}

According to the Hamiltonian of Eq. (\ref{b1}) the dark state of the free space can be expressed as

\begin{equation}\label{b2}
 |D\rangle=\frac{1}{\sqrt{2}}(|10\rangle|0\rangle -|01\rangle|0\rangle),
\end{equation}

\noindent and the bright state of the free space is

\begin{equation}\label{b3}
 |B\rangle=\frac{1}{\sqrt{2}}(|10\rangle|0\rangle +|01\rangle|0\rangle).
\end{equation}

\noindent Eqs. (\ref{b2}) and (\ref{b3}) are also known as antisymmetric and symmetric \cite{Rudolph 95, Ficek 14} states when not consider the cavity vacuum part (i.e. $|0\rangle$).

\renewcommand{\theequation}{C.\arabic{equation}}
\setcounter{equation}{0}  
\section*{APPENDIX C: THE MASTER EQUATIONS OF THE SYSTEM}  

We derive the equations of evolution for the elements of the density operator by using master equation of Eq. (\ref{15}) and the basis of Eq. (\ref{18}).

\begin{eqnarray}
\dot{\rho}_{ge_{1}}&=&i\omega_{0}\rho_{ge_{1}}-\gamma\rho_{ge_{1}}-(\gamma_{12}-i\Omega_{12})\rho_{ge_{2}}e^{-i\varphi} \notag \\&&
+i g_{1}\rho_{gc},
\end{eqnarray}
\begin{eqnarray}
\dot{\rho}_{ge_{2}}&=&i\omega_{0}\rho_{ge_{2}}-\gamma\rho_{ge_{2}}-(\gamma_{12}-i\Omega_{12})\rho_{ge_{1}}e^{i\varphi} \notag \\&&
+i g_{2}\rho_{gc}e^{i\varphi},
\end{eqnarray}
\begin{eqnarray}
\dot{\rho}_{gc}&=&i\omega_{c}\rho_{gc}+ig_{1}\rho_{ge_{1}}+ig_{2}\rho_{ge_{2}}e^{-i\varphi}-\kappa\rho_{gc},
\end{eqnarray}
\begin{eqnarray}
\dot{\rho}_{e_{1}e_{1}}&=&-2\gamma\rho_{e_{1}e_{1}}-\gamma_{12}(\rho_{e_{2}e_{1}}e^{i\varphi}+\rho_{e_{1}e_{2}}e^{-i\varphi})\notag\\&&
-i\Omega_{12}(\rho_{e_{2}e_{1}}e^{i\varphi}-\rho_{e_{1}e_{2}}e^{-i\varphi})-ig_{1}\rho_{ce_{1}} \notag\\&&
+ig_{1}\rho_{e_{1}c},
\end{eqnarray}
\begin{eqnarray}
\dot{\rho}_{e_{1}e_{2}}&=&-2\gamma\rho_{e_{1}e_{2}}-\gamma_{12}(\rho_{e_{1}e_{1}}e^{i\varphi}+\rho_{e_{2}e_{2}}e^{i\varphi})\notag\\&&
-i\Omega_{12}(\rho_{e_{2}e_{2}}e^{i\varphi}-\rho_{e_{1}e_{1}}e^{i\varphi})-ig_{1}\rho_{ce_{2}} \notag\\&&
+ig_{2}\rho_{e_{1}c}e^{i\varphi},
\end{eqnarray}
\begin{eqnarray}
\dot{\rho}_{e_{2}e_{2}}&=&-2\gamma\rho_{e_{2}e_{2}}-\gamma_{12}(\rho_{e_{2}e_{1}}e^{i\varphi}+\rho_{e_{1}e_{2}}e^{-i\varphi})\notag\\&&
-i\Omega_{12}(\rho_{e_{1}e_{2}}e^{-i\varphi}-\rho_{e_{2}e_{1}}e^{i\varphi})-ig_{2}\rho_{ce_{2}}e^{-i\varphi} \notag\\&&
+ig_{2}\rho_{e_{2}c}e^{i\varphi},
\end{eqnarray}
\begin{eqnarray}
\dot{\rho}_{e_{1}c}&=&-\gamma\rho_{e_{1}c}-\gamma_{12}\rho_{e_{2}c}e^{i\varphi}-\kappa\rho_{e_{1}c}-i\Omega_{12}\rho_{e_{2}c}e^{i\varphi}\notag\\&&
-ig_{1}\rho_{cc}+ig_{1}\rho_{e_{1}e_{1}}+ig_{2}\rho_{e_{1}e_{2}}e^{-i\varphi},
\end{eqnarray}
\begin{eqnarray}
\dot{\rho}_{e_{2}c}&=&-\gamma\rho_{e_{2}c}-\gamma_{12}\rho_{e_{1}c}e^{-i\varphi}-\kappa\rho_{e_{2}c}-i\Omega_{12}\rho_{e_{1}c}e^{-i\varphi}\notag\\&&
-ig_{2}(\rho_{cc}-\rho_{e_{2}e_{2}})e^{-i\varphi}+ig_{1}\rho_{e_{2}e_{1}},
\end{eqnarray}

\begin{eqnarray}
\dot{\rho}_{cc}&=&-ig_{1}(\rho_{e_{1}c}-\rho_{ce_{1}})-ig_{2}\rho_{e_{2}c}e^{i\varphi}\notag+ig_{2}\rho_{ce_{2}}e^{-i\varphi}\\&&
-2\kappa\rho_{cc},
\end{eqnarray}

\noindent where $\Delta=\omega_{0}-\omega_{c}$, $\varphi=\vec{k}_{0}\cdot(\vec{r}_{2}-\vec{r}_{1})=\frac{2\pi}{\lambda}r_{12}\cos\theta$, with $\theta$ being the angle between the propagating direction of driving field and the joining line of the two atoms. The position vectors of atoms can be any values because the dynamical behaviors of the system are only affected by the interatomic distance $r_{12}$. For simplicity, we here assume position vector $\vec{r_{1}}=0$, so only relative phase left in the density matrix equations.

\end{document}